\documentclass[comment,doublecol]{epl2}
\usepackage{bm}
\usepackage{amsmath}
\usepackage{amsthm}
\usepackage{amssymb}
\usepackage{mathtools}
\usepackage[unicode=true,
 bookmarks=false,
 breaklinks=false,pdfborder={0 0 1},backref=false,colorlinks=true]
 {hyperref}

\title{Comment on ``The Tully-Fisher law and dark matter effects derived
via modified symmetries'' by Arraut Ivan}
\shorttitle{Title} %Comment on ``The Tully-Fisher law and dark matter effects...'' by I. Arraut

\author{L.~Filipe~O.~Costa\inst{1} \and F. Frutos-Alfaro\inst{2} \and Jos\'{e} Nat\'{a}rio\inst{1} \and Michael Soffel\inst{3}}

\institute{                    
  \inst{1} CAMGSD, Departamento de Matemática, Instituto Superior Técnico, 1049-001
Lisboa, Portugal\\
  \inst{2} Space Research Center (CINESPA), School of Physics, University of
Costa Rica, 11501 San José, Costa Rica.\\
	\inst{3} Lohrmann Observatory, Dresden Technical University, 01062 Dresden,
Germany}

\begin{document}

\maketitle

\section{Introduction}
It has been claimed in \cite{Arraut2023} that General Relativity can explain the effects usually attributed to dark matter in galaxies, based on a metric where supposedly the angular momentum is replaced by the ``velocity'' as the conserved quantity associated to axial symmetry (described as a ``new modified symmetry that emerges at galactic scale''). This is claimed to lead to an `equal arcs in equal time'' orbital law, and to the MOND relation $v^{4}\propto Ma_{0}$, describing flat rotation curves.

We show all these claims to be false, resulting from an unphysical spacetime violating the energy conditions, combined with incorrect definitions of angular momentum, velocity, and arclength in the proposed geometry, plus a series of errors in the equations.
\section{Metric}

The proposed metric for describing the galactic gravitational field
is (cf. \cite{Arraut2023}, p. 3), in geometrized units $G=c=1$,
\begin{equation}
ds^{2}=-\left[1-\frac{2M}{r}\right]dt^{2}+\left[1-\frac{2M}{r}\right]^{-1}dr^{2}+rd\Omega^{2}\ ,\label{eq:ArrautMetric}
\end{equation}
which differs from the Schwarzschild metric in the angular part, having
$rd\Omega^{2}$ instead of $r^{2}d\Omega^{2}$ (implying $r$ and
$t$ to be dimensionless coordinates, $ds$ a dimensionless interval,
and $M$ to no longer be a mass). This is an \emph{unphysical metric},
totally \emph{unsuitable to describe galaxies}. It is not asymptotically
flat, and it is filled everywhere with exotic matter violating the strong
energy condition:
\begin{equation}
\mathbb{E}_{\ \alpha}^{\alpha}=8\pi\left[T_{\alpha\beta}-\frac{1}{2}T_{\ \gamma}^{\gamma}g_{\alpha\beta}\right]u^{\alpha}u^{\beta}=4\pi(\rho+\sigma)=-\frac{M}{r^3}<0\ .
\end{equation}
 Here $T^{\alpha\beta}$ is the energy-momentum tensor, $u^{\alpha}=\delta_{0}^{\alpha}(-g_{00})^{-1/2}$
is the 4-velocity of the observers at rest in the coordinate system
of Eq. \eqref{eq:ArrautMetric}, and $\mathbb{E}_{\alpha\beta}=R_{\alpha\mu\beta\nu}u^{\mu}u^{\nu}$,
$\rho\equiv T^{\alpha\beta}u_{\alpha}u_{\beta}=(r+4r^{2}-6M)/(32\pi r^{3})$,
and $\sigma=-(r+4r^{2}+2M)/(32\pi r^{3})$ are, respectively, the tidal
tensor, the energy density, and the trace of the spatial stresses $\sigma_{\alpha\beta}\equiv h_{\ \alpha}^{\mu}h_{\ \beta}^{\nu}T_{\mu\nu}$
($h_{\ \alpha}^{\mu}\equiv u^{\mu}u_{\alpha}+g_{\ \alpha}^{\mu}$)
as measured by these observers.
The weak (thus the dominant) energy condition is also violated, in limited regions, for certain values of the parameter $M$; e.g., for $M=0.01$, $\rho<0$ for the rest observers within $r<5M$, and within $r<14M$ for observers moving azimuthally at 99\% of the speed of light. 

The exotic matter has moreover negative Komar mass everywhere off
the singularity at $r=0$. The Komar mass $m=(1/8\pi)\int_{\partial\mathcal{V}}\star\mathbf{d}\bm{\xi}$
{[}where $(\star\mathbf{d}\bm{\xi})_{\alpha\beta}\equiv\xi_{\nu;\mu}\epsilon_{\ \ \alpha\beta}^{\mu\nu}$;
$\xi^{\alpha}=\delta_{0}^{\alpha}${]}, contained within a sphere
$r\le r_{{\rm S}}$, is $m(r_{{\rm S}})=M/r_{{\rm S}}$. It follows
that (i) the total Komar mass present in the spacetime is zero: $\lim_{r_{{\rm S}}\rightarrow\infty}m(r_{{\rm S}})=0$;
(ii) the singularity at $r=0$ has infinite positive Komar mass: $\lim_{r_{{\rm S}}\rightarrow0}m(r_{{\rm S}})=\infty$;
and (iii) the Komar mass inside any hollow sphere centered at $r=0$ is
negative.

It is instructive writing the metric \eqref{eq:ArrautMetric} in terms
of coordinates $\{X^{\alpha}\}$ with the dimensions of length. Any spherically symmetric
static metric can be written in the form
\begin{equation}
dS^{2}=g_{TT}(R)dT^{2}+g_{RR}(R)dR^{2}+R^{2}d\Omega^{2}\label{eq:SpheGen}
\end{equation}
where $R$, $T$ and $dS$ have dimensions of length, $R$ being the
areal radius. The metric \eqref{eq:ArrautMetric} can be written in
this form through the coordinate substitution $R=\mathcal{R}\sqrt{r}$,
$T=\mathcal{R}t$ (so that $dS^{2}=\mathcal{R}^{2}ds^{2}$), yielding
\begin{equation}
g_{TT}=-1+\frac{2\mathcal{M}^{2}}{R^{2}};\qquad g_{RR}=\left[1-\frac{2\mathcal{M}^{2}}{R^{2}}\right]^{-1}\frac{4R^{2}}{\mathcal{R}^{2}}\ ,\label{eq:MetricLength}
\end{equation}
where $\mathcal{R}$ and $\mathcal{M}\equiv\mathcal{R}\sqrt{M}$ are
constants with the dimensions of length (i.e., of mass). The 
non-asymptotic flatness of the metric is obvious from ${\rm lim}_{R\rightarrow\infty}g_{RR}=\infty$.

Observe finally that the dimensionless line element and coordinates in \eqref{eq:ArrautMetric} amount to measuring lengths in units of the length scale $\mathcal{R}$. 

\section{On the claimed new \textquotedblleft symmetries\textquotedblright\label{sec:On-the-claimed Symmetries}}

It is claimed that, in the metric \eqref{eq:ArrautMetric}, the quantity $\gamma$ in Eq. (10) of \cite{Arraut2023},
dubbed the ``velocity'', replaces the ``angular momentum'' as the conserved quantity associated
to axial symmetry; and that this leads to
an ``equal arcs in equal time'' orbital law (by contrast with the
Keplerian equal areas in equal time). These assertions are but \emph{misunderstandings},\emph{
stemming from incorrect definitions of arclength, velocity, and angular
momentum} in this geometry. Let $U^{\alpha}=dx^{\alpha}/d\lambda$
be the 4-velocity of a test particle in geodesic motion, assumed equatorial ($\theta=\pi/2$), where $\lambda$ is the proper time in units of $\mathcal{R}$, as implicitly used in \cite{Arraut2023}. The conserved quantity \cite{schutz_2009} $U_{\alpha}\partial_{\phi}^{\alpha}=U_{\phi}$ associated
to the Killing vector field $\partial_{\phi}^{\alpha}=\delta_{\phi}^{\alpha}$ in \eqref{eq:ArrautMetric} is indeed $\gamma$,
\begin{equation}
U_{\phi}=g_{\phi\phi}\frac{d\phi}{d\lambda}=r\frac{d\phi}{d\lambda}\equiv\gamma\ ;\label{eq:AngMomentum}
\end{equation}
however, this is not the magnitude of some spatial velocity. Recall, from Eqs. \eqref{eq:ArrautMetric} and  \eqref{eq:SpheGen}-\eqref{eq:MetricLength},
that the areal radius is $R=\sqrt{r} \mathcal{R}$, or, in units of $\mathcal{R}$, $\sqrt{r}$ (not $r$), so the infinitesimal
arc-length (in units of $\mathcal{R}$) along an equatorial circle is 
\begin{equation}
dl=\sqrt{dx^{i}dx^{j}g_{ij}}=\sqrt{g_{\phi\phi}}d\phi=\sqrt{r}d\phi .\label{eq:dl}
\end{equation}
The derivative of the arc-length $l$ with respect to the
coordinate time (i.e., the quantity sometimes called the speed, or ``linear
velocity'') is, for a circular equatorial motion in the metric \eqref{eq:ArrautMetric},
\begin{equation}
v=\frac{dl}{dt}=\sqrt{g_{\phi\phi}}\frac{d\phi}{dt}=\sqrt{r}\frac{d\phi}{dt}=R\frac{d\phi}{dT}\ . \label{eq:v}
\end{equation}
With respect to proper time, we have $dl/d\lambda=\sqrt{r}d\phi/d\lambda$,
differing from the quantity \eqref{eq:AngMomentum} by a factor $\sqrt{r}$.
It is thus clear that \eqref{eq:AngMomentum} is not a derivative
of arc-length with respect to proper time either, hence its conservation
does not yield an ``equal arcs in equal time'' law.

Equation \eqref{eq:AngMomentum} is actually precisely the quantity
known \cite{schutz_2009} as azimuthal angular momentum
per unit mass of the particle in this spacetime (measured in units of $\mathcal{R}$). This interpretation is evident in the coordinates with dimensions of length of \eqref{eq:SpheGen}, where we have $U_{\alpha}\partial_{\phi}^{\alpha}=R^{2} d\phi /d\tau$, with $U^{\alpha}=dX^{\alpha}/d\tau$ and $\tau=\mathcal{R}\lambda$ the proper time with dimensions of length. Observe that this holds for any spherically symmetric metric.

It should also be noted that the initial premise in \cite{Arraut2023}
that the observed galactic dynamics (supposedly the flat rotation
curves for circular orbits) reflects the conserved quantity along
a non-circular geodesic to be the ``velocity'' instead of the angular
momentum, is just another misconception, immediately negated by the
geodesic equation, regardless of the geometry considered.

\section{On the claim that this geometry reproduces the effects otherwise attributed
to dark matter}

It is claimed that the equation $v^{4}=4Ma_{0}$ is derived (Eq. (21) in
\cite{Arraut2023}, dubbed ``Tully-Fisher law in MONDIan language'', and actually differing from the MOND equation (2) of \cite{MOND_Galaxies} by a factor of 4). Such derivation is however false, consisting of a series
of errors. Equation (9) of \cite{Arraut2023}, claimed to follow from
substitution of Eq. (6) in Eq. (10) therein, is wrong {[}and inconsistent
with Eq. (10){]}; such substitution yields instead $L^{2}/r^{2}=\gamma^{2}$.
This is then used to obtain the wrong equations (16)-(18) in \cite{Arraut2023}.
A further misunderstanding is added from (18) of \cite{Arraut2023}
onward: in the inline expression $L=r \nu$ {(}$L\equiv r^{2}d\phi/d\lambda$, and $\nu$ stands here for the quantity denoted by $v$ in \cite{Arraut2023}{)}, the quantity $\nu=rd\phi/d\lambda\equiv\gamma$ is implied to be an orbital velocity;
as discussed in the previous Section, this is
false. Finally, another error is made in assuming that the quantity $\nu^2/r$
(incorrectly dubbed ``acceleration'') equals the MOND scale $a_{0}$; 
this is wrong (in several ways): $\nu^2/r$ is not an acceleration, but a squared velocity; the velocity of the
circular orbits tends to a constant in the MOND regime \cite{MOND_Galaxies,MOND_Galaxy_systems} (describing a flat velocity profile), as is well known; however such constant is $v=(Ma_{0})^{1/4}\ne a_{0}^{1/2}$, as stated by the MOND equation (2) of \cite{MOND_Galaxies} that the author aimed to derive.

The correct angular velocity $\Omega$ of circular geodesics in the metric \eqref{eq:ArrautMetric}
is readily obtained by writing the geodesic equation in the form \cite{schutz_2009}
$dU_{\alpha}/d\lambda=g_{\mu\nu,\alpha}U^{\mu}U^{\nu}/2$, whose $r$-component
yields, using $U^{\alpha}=U^{0}(\delta_{0}^{\alpha}+\Omega\delta_{\phi}^{\alpha})$,
\begin{align}
\Omega\equiv\frac{d\phi}{dt} & =\sqrt{\frac{-g_{00,r}}{g_{\phi\phi,r}}}=\frac{\sqrt{2M}}{r}\ ,\label{eq:OmegaGeo}
\end{align}
and the corresponding speed curve $v$, cf. Eq. \eqref{eq:v}, 
\begin{equation}
v=\frac{dl}{dt}=\sqrt{g_{\phi\phi}}\Omega=\sqrt{\frac{2M}{r}}=\sqrt{2}\frac{\mathcal{M}}{R}\ ,
\end{equation}
which is \emph{not a constant}. It is thus\emph{ false} that the geometry
described by \eqref{eq:ArrautMetric} yields flat rotation curves
such as those observed in galaxies and obtained in MOND, or that it
can describe galactic dynamics in any way.

\section{Conclusion}

We have shown all the main claims in Ref. \cite{Arraut2023}
to be false. The proposed geometry not only is unphysical, as it does
not even lead to the claimed MONDian equations, or mimics the dark matter effects in galaxies in any way; the conclusions to this effect in
\cite{Arraut2023} originate simply from errors in the equations.
The claims that in this geometry the velocity replaces the angular
momentum as the conserved quantity associated to the axial symmetry,
and that it leads to an ``equal arcs in equal time'' orbital law,
are also misunderstandings, stemming from incorrect definitions
of arclength, velocity and angular momentum. The initial premise
that the observed galactic dynamics reflects such properties is itself
also a misconception.%

\end{document}